\documentclass[
  aps,
  prl,
  twocolumn,
  10pt,
  superscriptaddress,
  floatfix,
  nofootinbib
]{revtex4-2}

\usepackage{amsmath,amssymb}
\usepackage{bm}            

\usepackage{siunitx}
\usepackage[section]{placeins}
\DeclareSIUnit\angstrom{\text{\AA}}

\usepackage{graphicx}
\usepackage{placeins}      
\usepackage{color}         
\usepackage[percent]{overpic}
\usepackage[normalem]{ulem}   

\usepackage[
  colorlinks=true,
  linkcolor=blue,
  citecolor=blue,
  urlcolor=blue
]{hyperref}

\makeatletter

\newcounter{smsec}

\newcounter{EMsection}
\renewcommand{\theEMsection}{EM\arabic{EMsection}}

\newcommand{\beginendmatter}{%
  \clearpage
  \setcounter{EMsection}{0}%
  \setcounter{figure}{0}%
  \setcounter{equation}{0}%
  \setcounter{table}{0}%
  \renewcommand{\thefigure}{EM\arabic{figure}}%
  \renewcommand{\theequation}{EM\arabic{equation}}%
  \renewcommand{\thetable}{EM\arabic{table}}%
}

\newcommand{\EMsection}[2]{%
  \refstepcounter{EMsection}%
  \phantomsection%
  \subsection*{\theEMsection.\ #1}%
  \label{#2}%
}
\makeatother

\usepackage{comment}

\begin{document}

\title{A Momentum-Resolved X-ray Thomson Scattering Benchmark of Electronic-Response Models in Warm Dense Aluminium}


\author{Dmitrii S. Bespalov}
\email{dmitrii.bespalov@xfel.eu}
\affiliation{European XFEL, Holzkoppel 4, 22869 Schenefeld, Germany}
\affiliation{Institute of Physics, University of Rostock, Albert-Einstein-Straße 23-24, 18059 Rostock, Germany}

\author{Ulf Zastrau}
\affiliation{European XFEL, Holzkoppel 4, 22869 Schenefeld, Germany}

\author{Zhandos A. Moldabekov}
\affiliation{Institute of Radiation Physics, Helmholtz-Zentrum Dresden-Rossendorf (HZDR), D-01328 Dresden, Germany}
\affiliation{Helmholtz-Zentrum Dresden-Rossendorf (HZDR), D-01328 Dresden, Germany}

\author{Thomas Gawne}
\affiliation{Center for Advanced Systems Understanding (CASUS), D-02826 G\"orlitz, Germany}
\affiliation{Helmholtz-Zentrum Dresden-Rossendorf (HZDR), D-01328 Dresden, Germany}

\author{Tobias Dornheim}
\affiliation{Institute of Radiation Physics, Helmholtz-Zentrum Dresden-Rossendorf (HZDR), D-01328 Dresden, Germany}
\affiliation{Center for Advanced Systems Understanding (CASUS), D-02826 G\"orlitz, Germany}
\affiliation{Helmholtz-Zentrum Dresden-Rossendorf (HZDR), D-01328 Dresden, Germany}

\author{Moyassar Meshhal}
\affiliation{Institute of Physics, University of Rostock, Albert-Einstein-Straße 23-24, 18059 Rostock, Germany}

\author{Alexis Amouretti}
\affiliation{Osaka University, Graduate School of Engineering Science, 1-3 Machikaneyama, Toyonaka, Osaka 560-8531, Japan}
\affiliation{European Synchrotron Radiation Facility (ESRF), 71 Avenue des Martyrs, CS 40220, 38043 Grenoble Cedex 9, France}

\author{Michal Andrzejewski}
\affiliation{European XFEL, Holzkoppel 4, 22869 Schenefeld, Germany}

\author{Karen Appel}
\affiliation{European XFEL, Holzkoppel 4, 22869 Schenefeld, Germany}

\author{Carsten Baehtz}
\affiliation{Helmholtz-Zentrum Dresden-Rossendorf (HZDR), D-01328 Dresden, Germany}

\author{Erik Brambrink}
\affiliation{European XFEL, Holzkoppel 4, 22869 Schenefeld, Germany}

\author{Khachiwan Buakor}
\affiliation{European XFEL, Holzkoppel 4, 22869 Schenefeld, Germany}

\author{Carolina Camarda}
\affiliation{European XFEL, Holzkoppel 4, 22869 Schenefeld, Germany}

\author{David Chin}
\affiliation{University of Rochester, Laboratory for Laser Energetics (LLE), 250 East River Road, Rochester, NY 14623, USA}

\author{Gilbert Collins}
\affiliation{University of Rochester, Laboratory for Laser Energetics (LLE), 250 East River Road, Rochester, NY 14623, USA}
\affiliation{University of Oxford, Department of Physics, Clarendon Laboratory, Parks Road, Oxford OX1 3PU, United Kingdom}

\author{C\'eline Cr\'episson}
\affiliation{University of Oxford, Department of Physics, Clarendon Laboratory, Parks Road, Oxford OX1 3PU, United Kingdom}

\author{Adrien Descamps}
\affiliation{Queen's University Belfast, School of Maths and Physics, University Road, Belfast BT7 1NN, United Kingdom}

\author{Jon Eggert}
\affiliation{Lawrence Livermore National Laboratory (LLNL), Physical \& Life Sciences, 7000 East Avenue, Livermore, CA 94550, USA}

\author{Luke Fletcher}
\affiliation{Stanford University and SLAC National Accelerator Laboratory, SIMES, Stanford, CA 94305, USA}

\author{Alessandro Forte}
\affiliation{University of Oxford, Department of Physics, Clarendon Laboratory, Parks Road, Oxford OX1 3PU, United Kingdom}

\author{Gianluca Gregori}
\affiliation{University of Oxford, Department of Physics, Clarendon Laboratory, Parks Road, Oxford OX1 3PU, United Kingdom}

\author{Marion Harmand}
\affiliation{CNRS -- DSI Meudon, Laboratoire PIMM (UMR 8006), ENSAM/CNAM, 155 Bd de l'H\^opital, 75013 Paris, France}

\author{Oliver S. Humphries}
\affiliation{European XFEL, Holzkoppel 4, 22869 Schenefeld, Germany}

\author{Hauke H\"oppner}
\affiliation{Helmholtz-Zentrum Dresden-Rossendorf (HZDR), D-01328 Dresden, Germany}

\author{Jonas Kuhlke}
\affiliation{Helmholtz-Zentrum Dresden-Rossendorf (HZDR), D-01328 Dresden, Germany}

\author{William Lynn}
\affiliation{Queen's University Belfast, School of Maths and Physics, University Road, Belfast BT7 1NN, United Kingdom}

\author{Julian L\"utgert}
\affiliation{Institute of Physics, University of Rostock, Albert-Einstein-Straße 23-24, 18059 Rostock, Germany}

\author{Masruri Masruri}
\affiliation{Helmholtz-Zentrum Dresden-Rossendorf (HZDR), D-01328 Dresden, Germany}

\author{Emma M. McBride}
\affiliation{Queen's University Belfast, School of Maths and Physics, University Road, Belfast BT7 1NN, United Kingdom}

\author{Ryan Stewart McWilliams}
\affiliation{The University of Edinburgh, School of Physics and Astronomy, JCMB, Peter Guthrie Tait Road, Edinburgh EH9 3FD, United Kingdom}

\author{Alan Augusto Sanjuan Mora}
\affiliation{Institute of Physics, University of Rostock, Albert-Einstein-Straße 23-24, 18059 Rostock, Germany}

\author{Jean-Paul Naedler}
\affiliation{Institute of Physics, University of Rostock, Albert-Einstein-Straße 23-24, 18059 Rostock, Germany}

\author{Paul Neumayer}
\affiliation{GSI Helmholtzzentrum f\"ur Schwerionenforschung, Postfach 11 05 52, 64278 Darmstadt, Germany}

\author{Charlotte Palmer}
\affiliation{Queen's University Belfast, School of Maths and Physics, University Road, Belfast BT7 1NN, United Kingdom}

\author{Alexander Pelka}
\affiliation{Helmholtz-Zentrum Dresden-Rossendorf (HZDR), D-01328 Dresden, Germany}

\author{Lea Pennacchioni}
\affiliation{Universit\"at Potsdam, Institut f\"ur Geowissenschaften, Karl-Liebknecht-Str. 24-25, 14476 Potsdam-Golm, Germany}

\author{Calum Prestwood}
\affiliation{European XFEL, Holzkoppel 4, 22869 Schenefeld, Germany}
\affiliation{Queen's University Belfast, School of Maths and Physics, University Road, Belfast BT7 1NN, United Kingdom}

\author{Natalia A. Pukhareva}
\affiliation{Institute for Materials, Ruhr-Universit\"at Bochum, Universit\"atsstr. 150, 44801 Bochum, Germany}

\author{Chongbing Qu}
\affiliation{Institute of Physics, University of Rostock, Albert-Einstein-Straße 23-24, 18059 Rostock, Germany}

\author{Divyanshu Ranjan}
\affiliation{DESY, Photon Science, Notkestrasse 85, 22607 Hamburg, Germany}
\affiliation{Institute of Physics, University of Rostock, Albert-Einstein-Straße 23-24, 18059 Rostock, Germany}

\author{Ronald Redmer}
\affiliation{Institute of Physics, University of Rostock, Albert-Einstein-Straße 23-24, 18059 Rostock, Germany}

\author{Michael R\"oper}
\affiliation{DESY, Photon Science, Notkestrasse 85, 22607 Hamburg, Germany}

\author{Christoph Sahle}
\affiliation{European Synchrotron Radiation Facility (ESRF), 71 Avenue des Martyrs, CS 40220, 38043 Grenoble Cedex 9, France}

\author{Samuel Schumacher}
\affiliation{Institute of Physics, University of Rostock, Albert-Einstein-Straße 23-24, 18059 Rostock, Germany}

\author{Jan-Patrick Schwinkendorf}
\affiliation{Helmholtz-Zentrum Dresden-Rossendorf (HZDR), D-01328 Dresden, Germany}

\author{Melanie J. Sieber}
\affiliation{Universit\"at Potsdam, Institut f\"ur Geowissenschaften, Karl-Liebknecht-Str. 24-25, 14476 Potsdam-Golm, Germany}

\author{Madison Singleton}
\affiliation{SLAC National Accelerator Laboratory, LCLS, 2575 Sand Hill Road, MS103, Menlo Park, CA 94025, USA}

\author{Ethan Smith}
\affiliation{University of Rochester, Laboratory for Laser Energetics (LLE), 250 East River Road, Rochester, NY 14623, USA}

\author{Christian Sternemann}
\affiliation{DELTA / Fakult\"at Physik, Technische Universit\"at Dortmund, Maria-Goeppert-Mayer-Str. 2, 44227 Dortmund, Germany}

\author{Thomas Stevens}
\affiliation{University of Oxford, Department of Physics, Clarendon Laboratory, Parks Road, Oxford OX1 3PU, United Kingdom}

\author{Michael Stevenson}
\affiliation{Institute of Physics, University of Rostock, Albert-Einstein-Straße 23-24, 18059 Rostock, Germany}

\author{Cornelius Strohm}
\affiliation{DESY, Photon Science, Notkestrasse 85, 22607 Hamburg, Germany}

\author{Minxue Tang}
\affiliation{DESY, Photon Science, Notkestrasse 85, 22607 Hamburg, Germany}

\author{Monika Toncian}
\affiliation{Helmholtz-Zentrum Dresden-Rossendorf (HZDR), D-01328 Dresden, Germany}

\author{Toma Toncian}
\affiliation{Helmholtz-Zentrum Dresden-Rossendorf (HZDR), D-01328 Dresden, Germany}

\author{Thomas Tschentscher}
\affiliation{European XFEL, Holzkoppel 4, 22869 Schenefeld, Germany}

\author{Sam M. Vinko}
\affiliation{University of Oxford, Department of Physics, Clarendon Laboratory, Parks Road, Oxford OX1 3PU, United Kingdom}

\author{Justin S. Wark}
\affiliation{University of Oxford, Department of Physics, Clarendon Laboratory, Parks Road, Oxford OX1 3PU, United Kingdom}

\author{Max Wilke}
\affiliation{Universit\"at Potsdam, Institut f\"ur Geowissenschaften, Karl-Liebknecht-Str. 24-25, 14476 Potsdam-Golm, Germany}

\author{Dominik Kraus}
\affiliation{Institute of Physics, University of Rostock, Albert-Einstein-Straße 23-24, 18059 Rostock, Germany}

\author{Thomas R. Preston}
\email{thomas.preston@xfel.eu}
\affiliation{European XFEL, Holzkoppel 4, 22869 Schenefeld, Germany}

\date{\today}

\begin{abstract}
The robust diagnosis of conditions generated in warm dense matter (WDM) experiments remains a persistent challenge. Here we describe the measurement of shock-compressed aluminium at $50$\,GPa with angle-resolved femtosecond x-ray Thomson scattering (XRTS) over a wide range of scattering wavevectors at the European XFEL. The measured plasmon dispersion and line shape show that the \textit{de facto} standard approach for analysing XRTS spectra, using uniform-electron-gas models, systematically overestimates the resonance energy by up to $8\,\mathrm{eV}$. We present an approach using \textit{ab initio} methods which agrees within the experimental uncertainty and demonstrates how accounting for shock-induced disorder in shock-compressed systems is critical for their understanding, and providing evidence that \textit{ab initio} treatments are required for reliable XRTS inference in warm dense aluminium.
\end{abstract}

\maketitle

Warm dense matter (WDM) bridges the regime between condensed matter and plasma, in which Coulomb coupling, partial ionization, quantum degeneracy, and structural disorder coexist~\cite{Kraus2025,vorberger2025roadmapwarmdensematter,wdm_book,new_POP,Dornheim_review}. These conditions are prevalent in giant planet interiors, brown dwarfs, and inertial-confinement-fusion (ICF) targets on the path to ignition~\cite{fortov_review,becker,Kritcher2020,SAUMON20221,Betti2016,Hurricane_RMP_2023,Ignition_PRL_2024,hu_ICF,drake2018high}. In WDM, equations of state, opacities, and transport properties are governed by the dynamic structure factor $S(k,\omega)$, which is often extracted from x-ray Thomson scattering (XRTS). For the present purpose, $S(k,\omega)$ is the spectrum of electron density fluctuations resolved in momentum and energy and therefore the quantity directly probed by the x-ray scattering signal. In practice, simulations and analyses of XRTS almost always rely on the Chihara-type decomposition of the dynamic structure factor into bound and free electron contributions~\cite{Chihara_JPhysF1987,Gregori_PRE_2003}.  The free electron contributions are typically computed using the uniform-electron-gas (UEG) model.
State-of-the-art \textit{ab initio} density functional theory (DFT) simulations~\cite{wdm_book} do not rely on this assumption; instead, their accuracy is determined by the choice of the electronic exchange--correlation (XC) functional and, in the case of time-dependent DFT (TDDFT)~\cite{ullrich2012time}, by the dynamic XC kernel or potential, which must be approximated in practice~\cite{ullrich2012time,Goerigk_PCCP_2017}.  

These models are routinely used to infer density and temperature from XRTS spectra in laboratory WDM experiments, yet they have rarely been corroborated by momentum-resolved density-response measurements under independently diagnosed thermodynamic conditions~\cite{dornheim2025unraveling,Bohme_PRL_2022}. Establishing such a controlled, momentum-resolved, high-quality benchmark for standard WDM models has therefore remained an open challenge.

Aluminium is a well-established benchmark material for testing these correlation models. At ambient conditions, its inelastic x-ray scattering and electron energy-loss spectra reveal clear deficiencies of the UEG model for the plasmon dispersion and spectral weight~\cite{schuelke1993,tischler2003,cazzaniga2011,Krane_JPF_1978,sprosser1989aluminum,Gawne2024xray}. Its simple valence structure and well-characterized equilibrium properties allow discrepancies between experiment and theory to be traced to specific many-body approximations rather than to chemical complexity or phase uncertainty. For example, studies on this material have clearly demonstrated the importance of non-linear Landau damping and deviations from simple Drude behaviour in the WDM  regime~\cite{witte2018prl,witte2019pop}. However, earlier XRTS measurements of x-ray-heated~\cite{sperling2015xfel} and shock-compressed aluminium~\cite{preston2019momentum}  were limited by restricted angular coverage, insufficient shot statistics, and substantial state uncertainties, preventing decisive discrimination between competing correlation models in WDM regime. 
Consequently, even for aluminium -- a material well suited for model validation -- UEG-based models and finite-temperature TDDFT approaches, despite their widespread use, have not yet been tested against momentum-resolved experiments under independently diagnosed conditions, i.e., without fitting WDM parameters in the XRTS calculations to achieve agreement with the experimental data.

In this Letter, we address deficiencies in widely used electronic response models by measuring the plasmon dispersion and line shape of shock-compressed aluminum.
Measurements are made at a pressure of approximately $50$~GPa, corresponding to a density range from $\rho = 3.75~\mathrm{g\,cm^{-3}}$ to $4.5~\mathrm{g\,cm^{-3}}$ and an electron temperature of $T \approx 0.6~\mathrm{eV}$. 
Angle-resolved femtosecond XRTS measurements were performed at the European XFEL over a wide range of scattering wavevectors, $k = 0.99$--$2.57~\si{\per\angstrom}$. This range spans the collective long-wavelength plasmon regime through to the particle-hole continuum, where electronic XC effects become significant. In this regime, UEG-based descriptions -- bare RPA and static local-field corrections (LFCs) -- predict plasmon peaks that are systematically overly dispersive and insufficiently damped, overshooting the measured resonance energy $E_{\mathrm{peak}}(k)$ by $1\text{--}8\,\mathrm{eV}$ and underestimating the high-loss spectral tail. In contrast, finite-temperature TDDFT based on DFT-MD ionic snapshots reproduces both the dispersion and the skewed, Landau-damped line shapes across the full $k$ range within experimental uncertainty (Supplemental Material~\cite{SM}, Sec.~S4). These measurements therefore provide, to our knowledge, the first momentum-resolved validation of finite-temperature TDDFT against XRTS in warm dense matter under independently diagnosed thermodynamic conditions, and they reveal systematic UEG deficiencies precisely in the density-temperature regime where such models are most commonly employed for XRTS inference.

\begin{figure}
  \centering
  \includegraphics[width=\linewidth]{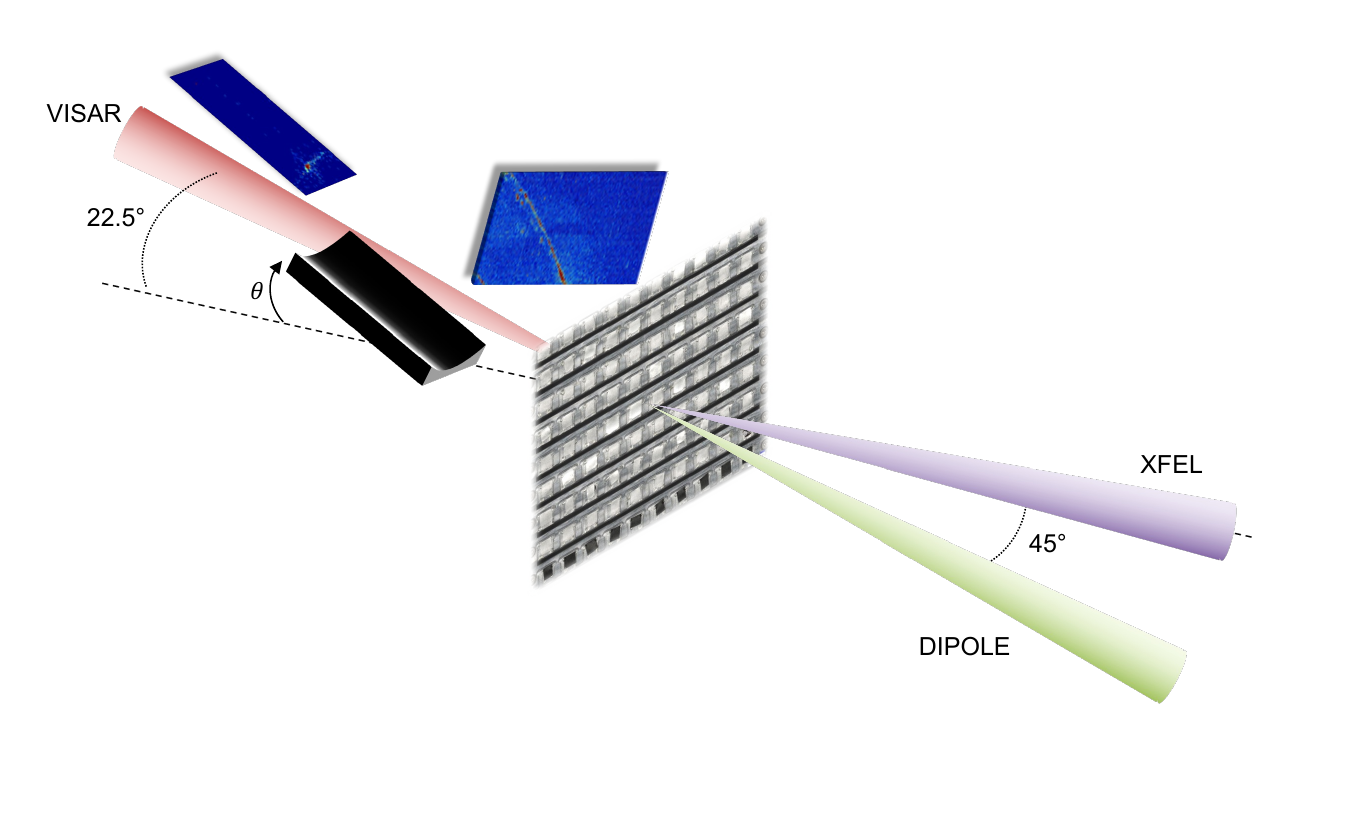}
  \caption{\textbf{Experimental geometry and diagnostics.}
The seeded XFEL probe (purple, $E_X\simeq8.3$~keV, $\Delta E\approx1$~eV) is focused onto the aluminium target by upstream Be compound refractive lenses (CRLs).
The target consists of a $50~\si{\micro\metre}$ Al foil bonded to a $25~\si{\micro\metre}$ Kapton ablator and is shock-compressed by a nanosecond DiPOLE drive laser (green) incident. Both beams are incident at $22.5^\circ$ relative to the target normal in the horizontal plane, and an angle between the beams of $45^\circ$.
Inelastically scattered x-rays are energy-dispersed by a cylindrically bent HAPG(004) crystal in von Hámos geometry and recorded on a Jungfrau detector, defining the scattering angle~$\Theta$ for XRTS in the vertical plane.
A second Jungfrau detector collects simultaneous x-ray diffraction (XRD), while velocity interferometer system for any reflector (VISAR, red) provides independent shock-timing and velocity diagnostics.}

  \label{fig:geom}
\end{figure}

Before examining the model comparison, we briefly summarize the experimental platform at the HED instrument~\cite{Zastrau2021} of the European XFEL sketched in Fig.~\ref{fig:geom}.  
A polycrystalline Al foil ($\sim50~\si{\micro\metre}$) bonded to a Kapton ablator ($\sim25~\si{\micro\metre}$) by a $\sim10~\si{\micro\metre}$ epoxy layer was probed by narrow-band self-seeded European XFEL pulses at $E_X=\SI{8307}{\eV}$ and driven by a frequency-doubled nanosecond DiPOLE laser \cite{Mason_2018_DiPOLE, butcher2015}.
Although the shocked target evolves on hydrodynamic timescales, the 25~fs XFEL probe is much shorter than that evolution, so each spectrum samples an effectively frozen state that can be treated within linear response.
Collective XRTS was recorded with a HAPG(004) von Hámos spectrometer~\cite{Preston2020} coupled to a Jungfrau detector~\cite{Mozzanica_2016}, while a second Jungfrau collected simultaneous wide-angle x-ray diffraction (XRD) (see End Matter, Sec.~\ref{sec:EM1}, and Supplemental Material~\cite{SM}, Sec.~S8).
Spectra were acquired at four scattering angles corresponding to $k=0.99$, $1.28$, $1.57$, and $2.57~\si{\per\angstrom}$, averaging $\sim40$ driven shots per geometry.
This combination of a femtosecond narrow-band FEL probe, a stable nanosecond drive, and concurrent XRTS/XRD diagnostics enables the momentum-resolved plasmon measurements and structural constraints required for a stringent benchmark of correlation models in warm dense aluminium.

\begin{figure*}[ht]
  \centering
  \includegraphics[height=8.5cm]{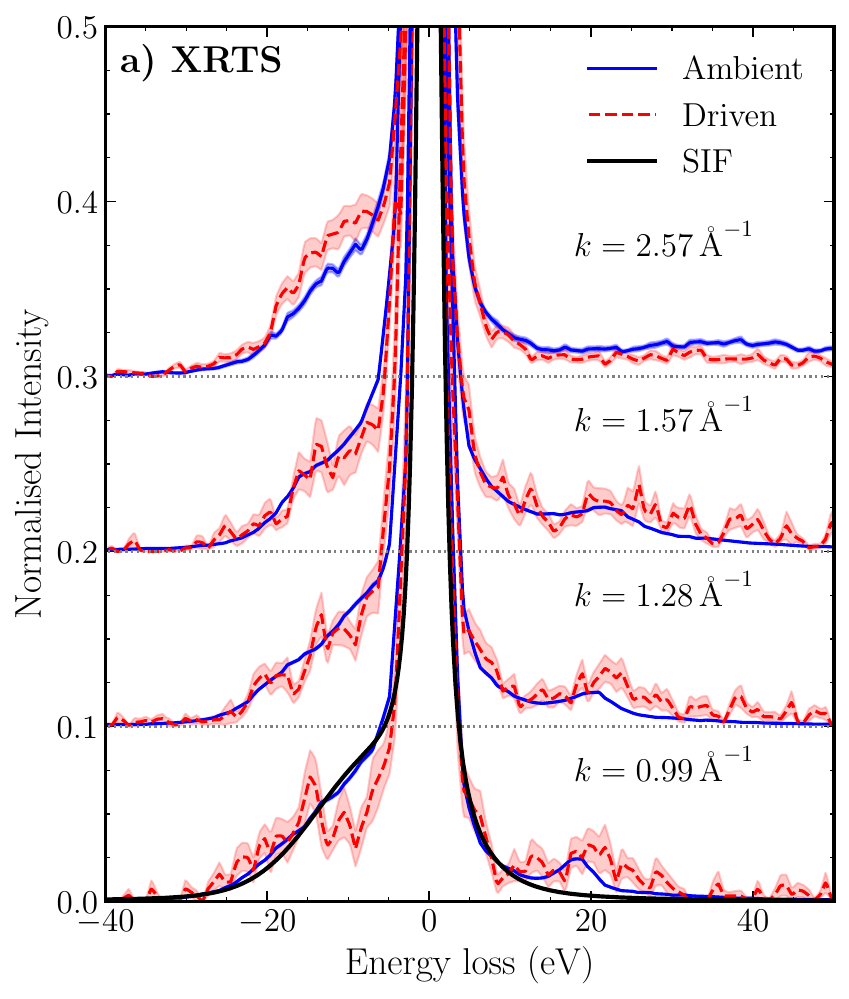}
  \includegraphics[height=8.5cm]{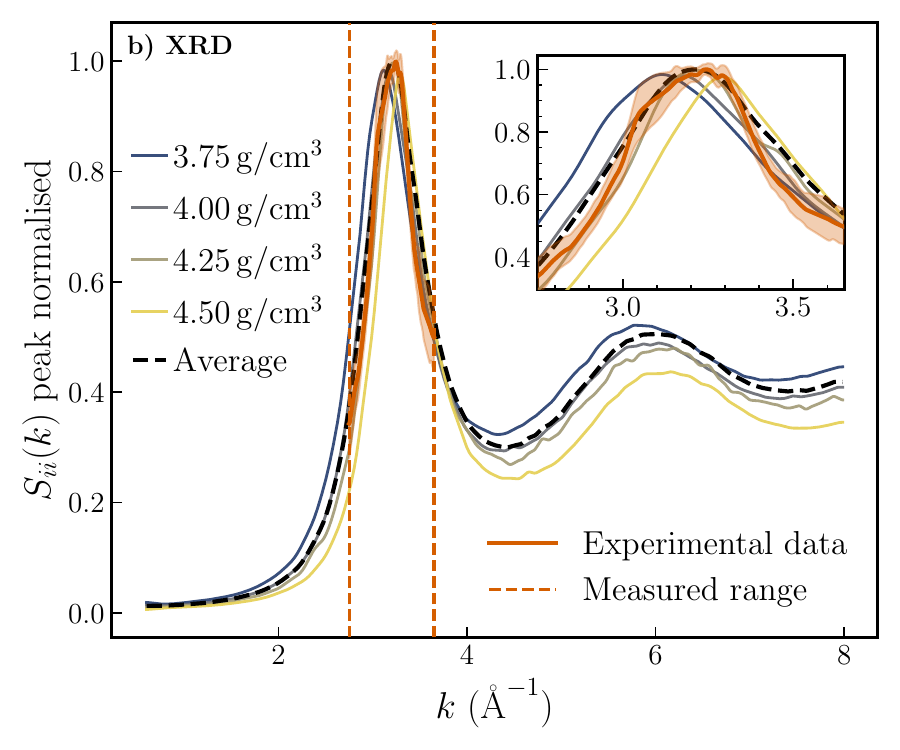}
 
  \caption{
  \textbf{Momentum-resolved XRTS and concurrent structural diagnostics in shock-compressed aluminium.}
  \textbf{(a)} Representative ambient (blue) and driven (red) spectra at four momentum transfers, normalised to the elastic line and vertically offset for clarity. The thin black curve shows the measured SIF at \(k=0.99~\si{\per\angstrom}\). The driven plasmon shifts to higher energy loss and broadens with increasing \(k\), becoming asymmetric near \(k=2.57~\si{\per\angstrom}\).
  \textbf{(b)} Ionic static structure factor \(S_{\mathrm{ii}}(k)\) extracted from XRD (orange line; band: $1\sigma$ shot-to-shot variation) compared with DFT-MD predictions at selected densities (coloured lines) and the density-averaged DFT-MD curve (black dashed) corresponding to the experimentally constrained density window. Vertical dashed lines indicate the experimental XRD \(k\)-range. The agreement constrains the density window used for theory comparisons.
  }
  \label{fig:XRTS_and_structure}
\end{figure*}

The compressed-state window is constrained by combining HELIOS-CR \cite{macfarlane2006helios} hydrodynamic simulations, VISAR \cite{10.1063/5.0271027} breakout timing, and simultaneous XRD (Fig.~\ref{fig:XRTS_and_structure}b; End Matter, Sec.~\ref{sec:EM1}, and Supplemental Material~\cite{SM}, Sec.~S8).
Together they indicate a $\sim\SI{30}{\micro\metre}$ plateau of shock-compressed aluminium at density, $\rho$, from $3.75~\mathrm{g\,cm^{-3}}$ to $4.5~\mathrm{g\,cm^{-3}}$ and nearly uniform temperature $T\simeq\SI{0.6}{\electronvolt}$ across the probed region.
The driven $S_{\mathrm{ii}}(k)$ extracted from XRD agrees with DFT-MD predictions within this density window (Fig.~\ref{fig:XRTS_and_structure}b and Supplemental Material~\cite{SM}, Sec.~S8), independently supporting the inferred state.  
In all model comparisons we therefore evaluate $S(k,\omega)$ at several conditions spanning the density range at $T\simeq\SI{0.6}{\electronvolt}$ and average the resulting spectra before convolution with the source-and-instrument function (SIF); this density averaging is applied consistently to all models. Within this framework, the role of the self-seeded XFEL coherence is to define the measured source bandwidth and instrumental response entering the SIF, rather than to modify the underlying many-body density response itself. The experimental comparison therefore tests the calculated $S(k,\omega)$ after convolution with the measured SIF under the independently constrained thermodynamic conditions described above.

Representative ambient and driven spectra at the four scattering vectors are shown in Fig.~\ref{fig:XRTS_and_structure}a. In the unshocked foil the inelastic response is dominated by a narrow plasmon peak clearly separated from the elastic line by the measured SIF, providing a check of the instrumental resolution and absolute energy calibration.  
In the shocked state the plasmon shifts to higher energy loss and broadens systematically with increasing $k$, developing a pronounced asymmetric high-loss wing as it approaches the particle-hole continuum~\cite{siegfried_review}, signalling the gradual loss of a long-wavelength collective mode and the growing influence of single-particle, Landau-damped excitations. 
For each scattering geometry, the SIF is determined from dedicated measurements on cold (ambient) aluminium and held fixed in the analysis (End Matter, Sec.~\ref{sec:EM2}; Supplemental Material~\cite{SM}, Sec.~S2), reflecting small geometry-dependent variations in the spectrometer response. The driven spectra have sufficient signal-to-noise that the plasmon peak position and the overall evolution of the line shape with $k$ are resolved. The residual fluctuations are noticeably smaller than the inelastic signal itself and are consistent with counting statistics superimposed on weak continuum and residual ablator backgrounds. The impact of the ablator is discussed in Supplemental Material~\cite{SM}, Sec.~S6.

\begin{figure*}[ht]
  \centering
  \includegraphics[width=0.49\linewidth]{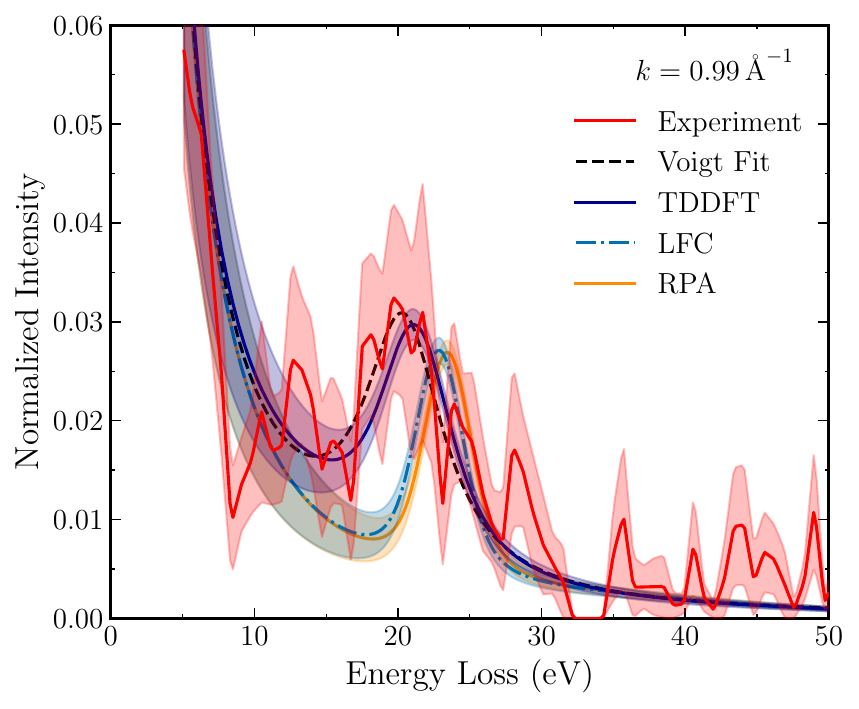}\hfill
  \includegraphics[width=0.49\linewidth]{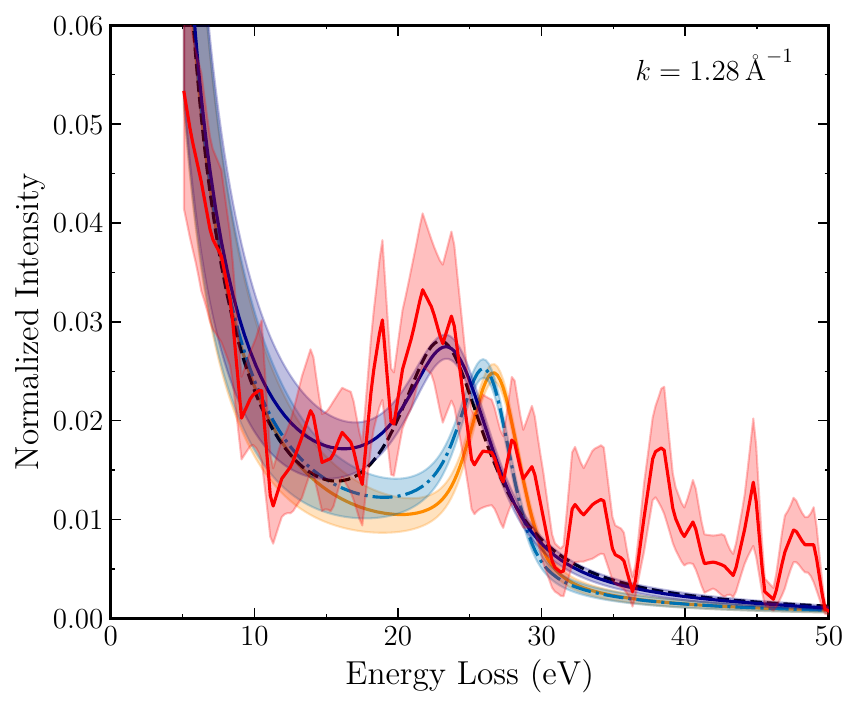}
  \caption{
  \textbf{Low-$k$ XRTS spectra and model comparison.}
  \textbf{(a)} $k=0.99~\si{\per\angstrom}$; \textbf{(b)} $k=1.28~\si{\per\angstrom}$.  
  Inelastic x-ray scattering spectra from laser-compressed aluminium at two representative momentum transfers, normalised to the elastic peak.  
  Experimental data (red lines, shaded band: $1\sigma$ uncertainty) are compared with theoretical predictions from TDDFT (blue), RPA (orange), and static LFCs (light blue).  
  A Voigt-profile plasmon (black dashed) is shown for reference and is used to extract the experimental plasmon peaks.  
  All theoretical spectra and the Voigt model are convolved with the experimentally determined SIF, modelled as the sum of two Voigt profiles (seed spike plus weak pedestal; End Matter, Sec.~\ref{sec:EM2}; Supplemental Material~\cite{SM}, Sec.~S2).  
  Only two $k$ values are shown here for clarity; spectra for all four measured $k$ are provided in Supplemental Fig.~S3.}
  \label{fig:comparison_spectra}
\end{figure*}

To extract plasmon energies and linewidths, we fit each ambient and driven spectrum with a minimal, physically motivated model applied identically at all scattering vectors.
The model consists of an elastic component fixed by the measured SIF, a Voigt-shaped plasmon contribution and a smooth background accounting for residual continuum and ablator scattering (End Matter, Sec.~\ref{sec:EM2}; see also Supplemental Material~\cite{SM}, Sec.~S2, and Fig.~\ref{fig:comparison_spectra}).
Statistical uncertainties from photon counting are propagated through the fit covariance, while systematic contributions from SIF calibration, absolute energy-scale drift, and scattering-angle uncertainty are combined in quadrature to obtain conservative error bars on the plasmon peak position $E_{\mathrm{peak}}(k)$ and width (End Matter, Sec.~\ref{sec:EM2}, and Supplemental Material~\cite{SM}, Sec.~S5).
At the largest measured momentum transfer, $k=2.57~\si{\per\angstrom}$, the inelastic feature in the scattering spectrum appears as a flat plateau, but the plasmon feature remains nevertheless identifiable as a clear maximum in the energy-loss spectrum and therefore still provides a robust determination of $E_{\mathrm{peak}}(k)$.
The resulting uncertainties are substantially smaller than those in earlier XRTS studies of warm dense aluminium~\cite{sperling2015xfel,preston2019momentum,witte2018prl} and remain well below the several-\si{\electronvolt} separation between UEG-based and TDDFT predictions (up to $\sim8~\si{\electronvolt}$ at $k=2.57~\si{\per\angstrom}$), ensuring that the model differences are cleanly resolved.

\begin{figure}
  \centering
  \includegraphics[width=\linewidth]{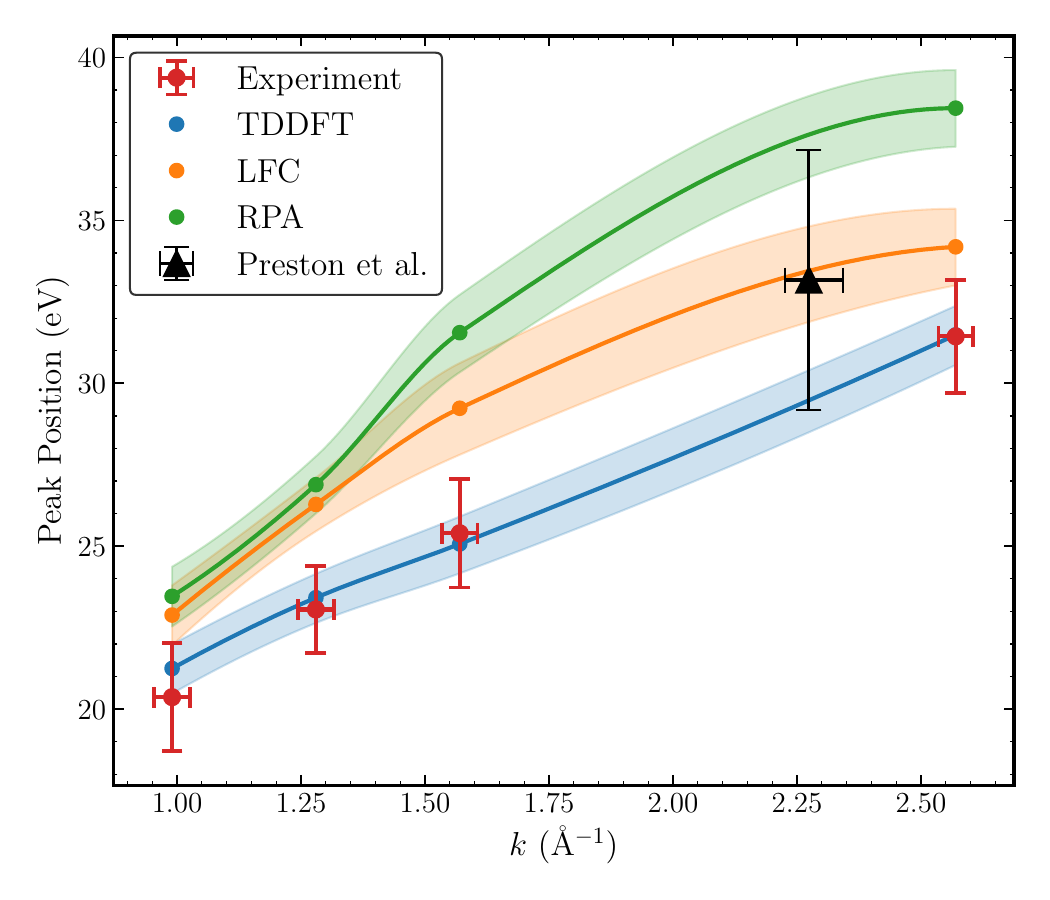}
  \caption{
  \textbf{Plasmon dispersion in warm dense aluminium and comparison to theory.}
  Extracted plasmon peak positions $E_{\mathrm{peak}}(k)$ (red markers) compared with theoretical predictions from TDDFT (blue line), static LFCs (orange), and RPA (green).  
  For reference, a representative data point from Preston \textit{et al.}~\cite{preston2019momentum} (black marker) illustrates the much larger experimental uncertainty typical of previous measurements at similar conditions.  
  Error bars include both fitting uncertainties and shot-to-shot experimental variation.  
  Lines connecting the data points are guides to the eye only.  
  TDDFT tracks the experimental dispersion within error bars at all measured $k$, whereas UEG RPA and static LFCs systematically overshoot the plasmon energy.}
  \label{fig:dispersion}
\end{figure}

Figure~\ref{fig:dispersion} compiles the extracted plasmon peak energies $E_{\mathrm{peak}}(k)$ and compares them with three classes of models, each evaluated over the HELIOS-CR/XRD-constrained density window and convolved with the measured SIF (End Matter, Secs.~\ref{sec:EM1} and \ref{sec:EM2}; full spectra and model curves are shown in Supplemental Material~\cite{SM}, Sec.~S4).  
UEG-RPA and its quantum-Monte-Carlo-based static local-field corrections systematically overestimate the plasmon energy and underpredict the high-loss spectral weight across the full momentum range, remaining overly dispersive and insufficiently damped even as the mode approaches the particle-hole continuum at large $k$~\cite{Fortmann_PRE_2010,Dornheim_PRL_2020_ESA,quantum_theory}.  
A weak contribution from the Kapton ablator affects only the far high-loss wing; ambient ablator-only measurements and spectrometer ray-tracing bound its impact and show that it does not shift the extracted plasmon peak positions or widths (Supplemental Material~\cite{SM}, Sec.~S6).  
In contrast, TDDFT based on finite-temperature DFT-MD ionic snapshots reproduces both the dispersion and the asymmetric, broadened line shapes within the experimental error bars.  
Quantitatively, UEG-based models overshoot $E_{\mathrm{peak}}(k)$ by up to $\Delta E\simeq8~\si{\electronvolt}$ at $k=2.57~\si{\per\angstrom}$, whereas TDDFT remains consistent with the data at all $k$ within the combined experimental and state-averaging uncertainties.  
The combination of high signal-to-noise spectra and error bars that are small compared with the $1\text{--}8~\si{\electronvolt}$ separation between UEG and TDDFT predictions makes this discrepancy a robust, model-discriminating test of electronic correlation treatments in warm dense aluminium.

This systematic breakdown of UEG-based approaches reflects the underlying real-space structure of the compressed liquid.
Finite-temperature DFT-MD (Supplemental Material~\cite{SM}, Sec.~S3) shows that driving aluminium into the experimentally constrained density window at an electron temperature of $T \approx 0.6~\si{\electronvolt}$ erases long-range crystalline order and localizes valence electrons into distorted coordination shells around the ions.
The resulting liquid exhibits strong ion--electron correlations and pronounced short-range structure in $S_{\mathrm{ii}}(k)$ and $g(r)$ that persist despite electronic degeneracy~\cite{Dornheim_review,dornheim_dynamic}.  
UEG models, even when supplemented by static LFCs, approximate this environment by a homogeneous electron liquid with averaged short-range corrections and cannot capture the $k$-dependent scattering phase shifts, collision-induced broadening, and coupling to the ionic structure that shape the measured response.  
TDDFT, by sampling the linear response on top of realistic ionic configurations and employing an effective dynamic exchange-correlation kernel, automatically incorporates these inhomogeneities and multiple-scattering pathways~\cite{ullrich2012time,Goerigk_PCCP_2017}.  
The close agreement between TDDFT and experiment therefore indicates that the plasmon dispersion and damping in warm dense aluminium are governed by this disordered, strongly correlated local environment rather than by small perturbations of a uniform-electron-gas state.

Our measurements provide clear experimental evidence that UEG-based Chihara-type models~\cite{Chihara_JPhysF1987,Gregori_PRE_2003} break down even for aluminium, long regarded as a prototypical nearly free-electron metal, when driven into the warm dense regime under independently diagnosed conditions. Because such models are widely used to infer density and temperature from XRTS spectra in high-energy-density experiments~\cite{Kraus2025, Glenzer_PRL_2007,vorberger2025roadmapwarmdensematter}, our results demonstrate that their application to warm dense aluminium at the probed densities and temperatures leads to a systematic, model-dependent bias in the inferred thermodynamic conditions and derived transport properties. \textit{Ab initio} approaches, such as finite-temperature time-dependent density-functional theory (TDDFT), should therefore be preferred wherever computationally feasible for quantitative XRTS interpretation under comparable warm dense matter conditions~\cite{Betti2016,Hurricane_RMP_2023,Ignition_PRL_2024}.

Together with the independently constrained thermodynamic state and full momentum coverage, the present data set -- including tabulated peak positions and complete spectral comparisons provided in the Supplemental Material~\cite{SM} (Sec.~S4) -- constitutes, to our knowledge, the first momentum-resolved benchmark for testing and developing dynamic exchange-correlation kernels and data-driven functionals in warm dense matter~\cite{Dornheim_PRL_2020_ESA,dornheim2025unraveling}. The European XFEL platform demonstrated here can now be extended to other elements, higher temperatures, and time-resolved shock sequences, enabling systematic
precision validation of state-of-the-art 
many-body theories under extreme conditions.

\begin{acknowledgments}
\textbf{Acknowledgements.} We acknowledge European XFEL for provision of beam time at the HED-HiBEF instrument under Proposal No.~6656 and thank the staff for their assistance. The authors are grateful to the HIBEF user consortium for the provision of instrumentation and staff that enabled this experiment. We thank S.~H.~Glenzer for helpful discussions. 

This work was funded by the Deutsche Forschungsgemeinschaft (DFG) under Grants RE~882/24-1 and ZA~828/4-1 (Project No.~493108501). K.B. acknowledges DFG Project AP262/2-2 (Project No.~280637173). C.C. acknowledges DFG Projects AP262/3-1 and STE1079/10-1 (Project No.~521549147). K.A., K.B., R.R., and T.T. acknowledge support within DFG Research Unit FOR~2440. L.B.F. acknowledges support by U.S.\ DOE Fusion Energy Sciences under FWP~100182. This work was partially supported by the Center for Advanced Systems Understanding (CASUS), financed by the German Federal Ministry of Education and Research and the Free State of Saxony. This project received funding from the European Research Council (ERC) under the European Union’s Horizon~2022 programme (Grant Agreement No.~101076233, ``PREXTREME''). It also received funding from the European Union’s Just Transition Fund (JTF) within the project \emph{R\"ontgenlaser-Optimierung der Laserfusion} (ROLF), Contract No.~5086999001, co-financed by the Saxon state government. Computations were performed at the Center for Information Services and High-Performance Computing (ZIH), TU Dresden, and at the Norddeutscher Verbund f\"ur Hoch- und H\"ochstleistungsrechnen (HLRN) under Grant mvp00024. We acknowledge the ``Zentrum f\"ur \mbox{Grenzfl\"achendominierte}
H\"ochstleistungswerkstoffe'' (ZGH) at Ruhr University Bochum for use of its XRD diffractometer. E.E.M.\ and A.D.\ were supported by the UKRI Future Leaders Fellowship (Grant No.~MR/W008211/1).

This material is based upon work supported by the Department of Energy [National Nuclear Security Administration] University of Rochester ``National Inertial Confinement Fusion Program'' under Award Number(s) DE-NA0004144. This report was prepared as an account of work sponsored by an agency of the United States Government. Neither the United States Government nor any agency thereof, nor any of their employees, makes any warranty, express or implied, or assumes any legal liability or responsibility for the accuracy, completeness, or usefulness of any information, apparatus, product, or process disclosed, or represents that its use would not infringe privately owned rights. Reference herein to any specific commercial product, process, or service by trade name, trademark, manufacturer, or otherwise does not necessarily constitute or imply its endorsement, recommendation, or favoring by the United States Government or any agency thereof. The views and opinions of authors expressed herein do not necessarily state or reflect those of the United States Government or any agency thereof.
\end{acknowledgments}

D.S.B. prepared targets, performed the experiment, analyzed the data, and drafted the manuscript with U.Z., T.G., and T.R.P.. T.R.P., Z.A.M., and T.G. contributed to analysis and writing. O.S.H., D.K., and T.R.P. supervised and contributed to experimental design, preparation, and analysis. K.A., M.A., C.B., E.B., O.S.H., H.H., A.P., J.-P.S., C.S., M.T. and T.R.P. set up the experiment; O.S.H. established online data processing. T.G. performed ray-tracing simulations. M.\,Masruri, M.\,Toncian, and T.\,Toncian operated the DiPOLE laser. E.B.\ and A.D. calibrated the VISAR. M.R. prepared the ablators. Z.A.M. performed TDDFT, DFT-MD, RPA, and LFC simulations; M.\,Meshhal performed DFT-MD; both supported theoretical analysis. N.A.P. recorded the XRD reference spectrum. All authors discussed the results and edited the manuscript.

Data availability: The data that support the findings of this article are available in the European XFEL repository~\cite{XFELdata6656}. An embargo may apply in accordance with facility policy; until release, the data are available from the authors upon reasonable request.

\nocite{Carnimeo_JCTC_2023,GawneMosaic,Giannozzi_2017,Giannozzi_jcp_2020,HEART,PP_ESPRESSO_2022,TIMROV2015460}
\bibliographystyle{apsrev4-2}
\bibliography{references_PRL}

\onecolumngrid

\beginendmatter


\EMsection{Experimental state determination}{sec:EM1}

The thermodynamic state of the shocked aluminium was constrained using VISAR breakout timing together with HELIOS-CR hydrodynamic simulations employing PROPACEOS equation-of-state and opacity tables~\cite{macfarlane2006helios,More1988QEOS,White2012GraphitePROPACEOS}.

Target compression was modelled with the 1D Lagrangian radiation-hydrodynamics code HELIOS-CR using the full Kapton/epoxy/Al target stack and the experimental laser parameters. The incident drive was represented as an 8~ns flat-top pulse at 515~nm with a nominal energy of $\sim26$~J. The absorbed drive energy was calibrated by matching the simulated breakout time to the optical breakout measured by VISAR, yielding an effective absorbed energy of $\sim19$~J.

Across the shot set the breakout time remained stable at $t_\mathrm{b}=6.5\pm0.1$~ns, while the XFEL probe occurred at $t_\mathrm{pp}=6.3$~ns, strictly before breakout. Profiles extracted at the probe time predict a $\sim30~\mu$m plateau of compressed aluminium with density $\rho=3.75$--$4.5~\mathrm{g\,cm^{-3}}$ and an approximately uniform temperature $T\approx0.6~\si{\electronvolt}$. Independent analysis of the driven XRD signal yields an ionic structure factor consistent with DFT-MD predictions in the same density window, supporting the state assignment used in the theory comparisons.

\begin{figure}[t]
\centering
\includegraphics[width=0.98\linewidth]{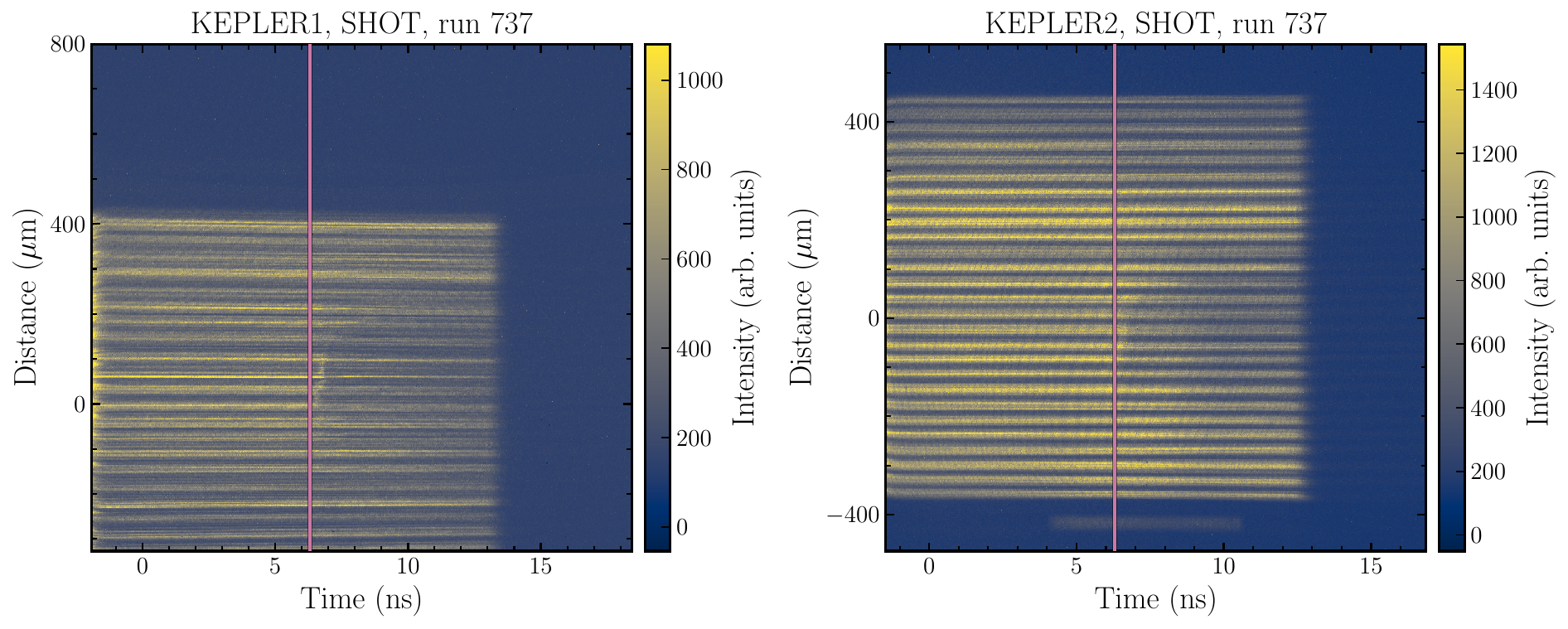}
\caption{\textbf{Representative VISAR shot used for timing calibration.}
Single-shot VISAR images from run 737 recorded on the two KEPLER channels. The vertical magenta line marks the XFEL probe time $t_\mathrm{pp}=6.3$~ns. The breakout time remains stable across the shot set at $t_\mathrm{b}=6.5\pm0.1$~ns, providing the timing constraint used to calibrate the HELIOS-CR simulations.}
\label{fig:EM_VISAR}
\end{figure}

\begin{figure}[ht]
\centering
\includegraphics[width=0.5\linewidth]{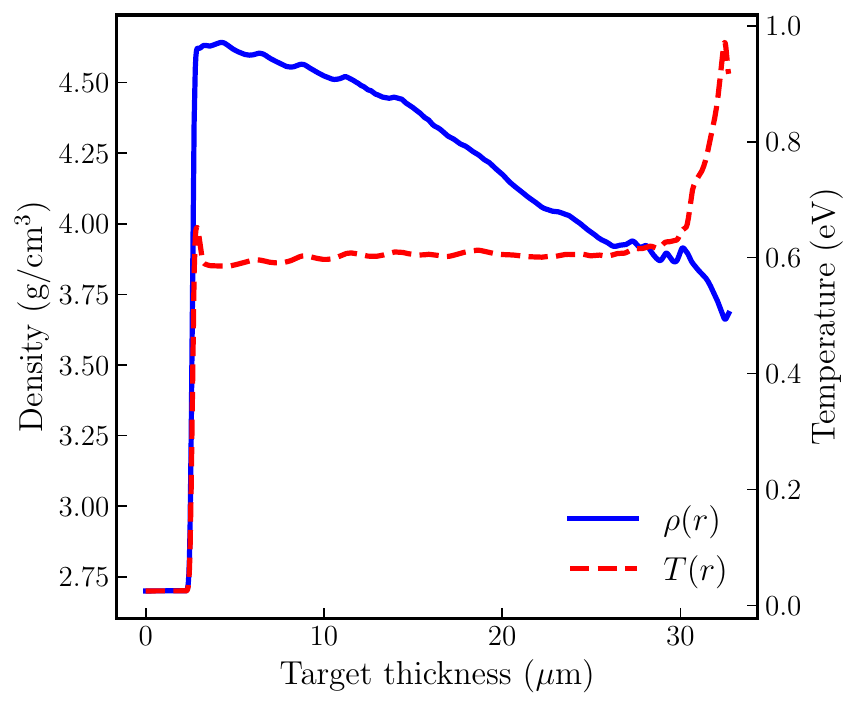}
\caption{\textbf{Hydrodynamic profiles at the XFEL probe time.}
Density (solid blue, left axis) and temperature (dashed red, right axis) from HELIOS-CR at $t_\mathrm{pp}=6.3$~ns. The simulation predicts a $\sim30~\mu$m-wide plateau of compressed aluminium with density $3.75$--$4.5~\mathrm{g\,cm^{-3}}$ and nearly constant $T\approx0.6$~eV. This density window is used for the density-averaged theory comparison in the Letter.}
\label{fig:EM_HydroProfile}
\end{figure}

\EMsection{Instrumental response and spectral analysis}{sec:EM2}

All experiment-theory comparisons use a measured source-and-instrument function (SIF) determined independently for each scattering geometry from cold aluminium targets acquired under identical spectrometer settings.

For each momentum transfer the quasi-elastic line was fitted with a two-Voigt model consisting of a narrow seeded component (FWHM $\approx3.3$~eV) and a weak pedestal (FWHM $\approx22.7$~eV), the latter fitted on the energy gain side where there is no inelastic scattering. The resulting effective spectral resolution is approximately $3.5$~eV (FWHM). These geometry-specific SIFs are then held fixed in all driven-shot fits and in all convolutions of the theoretical spectra.

During the experiment no measurable drift of the incident x-ray energy was observed on the scale relevant for the reported plasmon positions. The SIF fits were constrained such that the elastic peak remained centred at zero energy transfer, ensuring a stable absolute energy reference throughout the dataset.

For quantitative analysis the elastic contribution is fixed by the measured SIF while the inelastic plasmon feature is fitted with a Voigt profile. All theoretical spectra are convolved with the same geometry-specific SIF prior to comparison with the experimental data.

\begin{figure}[ht]
\centering
\includegraphics[width=0.5\linewidth]{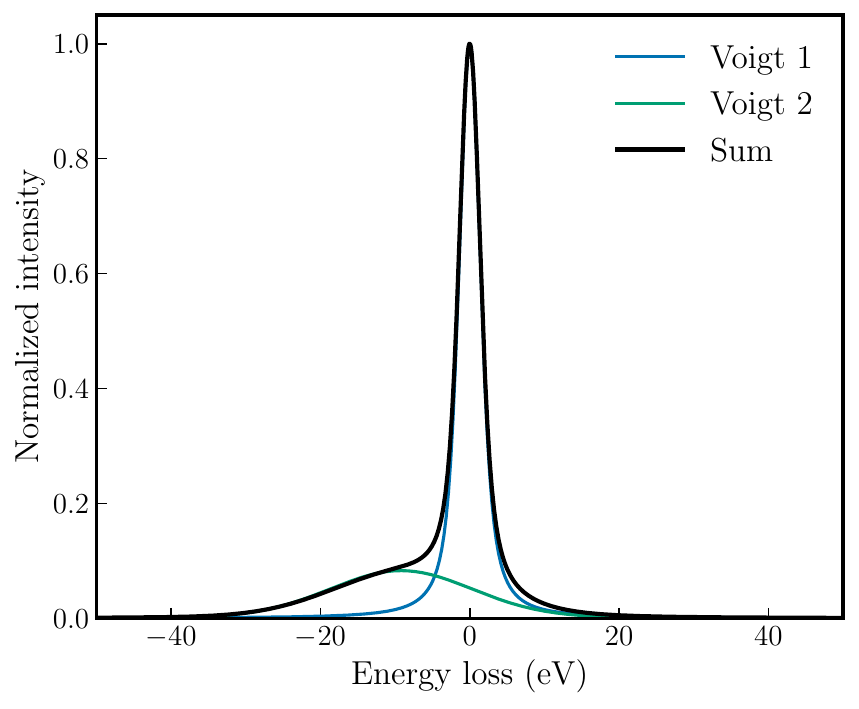}
\caption{\textbf{Representative source-and-instrument function for lowest k value.}
Two-Voigt decomposition of the elastic scattering signal measured from cold aluminium. The narrow component represents the seeded XFEL spike while the broader pedestal corresponds to the residual SASE background. All theoretical spectra are convolved with the measured SIF before comparison to experiment.}
\label{fig:EM_SIF}
\end{figure}

\clearpage

\end{document}